\documentstyle[aps,prb]{revtex}

\newcommand{\beq}{\begin{equation}}
\newcommand{\eeq}{\end{equation}}

\begin{document}

\twocolumn[\hsize\textwidth\columnwidth\hsize\csname
@twocolumnfalse\endcsname

\title{Energy loss of charged particles interacting with simple metal surfaces}
\author{A. Garc{\'\i}a-Lekue$^{1}$ and J. M. Pitarke$^{1,2}$}
\address{$^1$Materia Kondentsatuaren Fisika Saila, Zientzi Fakultatea,
Euskal Herriko Unibertsitatea,\\ 644 Posta kutxatila, 48080 Bilbo, Basque
Country, Spain\\
$^2$Donostia International Physics Center (DIPC) and Centro Mixto 
CSIC-UPV/EHU,\\
Basque Country, Spain}

\date{4 January 2001}

\maketitle

\begin{abstract}
Self-consistent calculations of the energy-loss spectra of charged particles
moving near a plane-bounded free electron gas are reported. Energy-loss
probabilities are obtained, within linear-response theory, from the knowledge
of the density-response function of the inhomogeneous electron system.
Self-consistent single-particle wave functions and energies are obtained by
solving the Kohn-Sham equation of density-functional theory, and the electronic
response is then computed either in the random-phase approximation or with the
use of an adiabatic local-density approximation. Special emphasis is placed on
the various contributions from collective and electron-hole excitations to the
energy loss of charged particles moving parallel with the surface. The effect
of the electronic selvage at a metal surface on the energy-loss spectra is also
discussed, by comparing our full self-consistent calculations with those
obtained for electron densities that drop abruptly to zero at the surface.
\end{abstract}

]

\section{Introduction}

Charged particles interacting with metal surfaces create electron-hole pairs
and, by virtue of the dynamically screened Coulomb interaction, bulk
and surface collective excitations, i.e., plasmons.\cite{Pines,Ritchie1} These
excitations play a crucial role in the interpretation of surface electron
spectroscopy data, such as x-ray photoelectron spectroscopy (XPS),
Auger-electron spectroscopy (AES), and reflection-electron energy-loss
spectroscopy (REELS).\cite{Raether} The interaction of swift electrons with
surfaces has also attracted great interest in the field of scanning transmission
electron microscopy (STEM).\cite{Batson,Kohl,Ritchie2,Howie} Equally, the
interaction of moving ions with solids has represented an active field of basic
and applied physics,\cite{exp1,exp2} and a great amount of research has recently
been focused on the case of ions that are incident at grazing
angle.\cite{exp2p,exp3,exp4,exp5} Nevertheless, existing calculations of
energy-loss spectra invoke either the local-dielectric, the hydrodynamic, or the
specular-reflexion model of the
surface.\cite{Pendry,Nunez,Zabala,Yubero,Chen,Juaristi,Ding,Ritchie3} An
exception is a recent self-consistent calculation of the stopping power of
jellium planar surfaces for ions moving parallel with the
surface.\cite{Cazalilla}

In this paper, we present extensive self-consistent calculations of the
energy-loss spectra of charged particles moving near a jellium surface. In
the case of charged particles moving inside a solid, nonlinear effects
are known to be crucial in the interpretation of energy-loss
measurements;\cite{Pitarke1,Pitarke2} however, these corrections have been
shown to be less important when the charged particle moves outside the
solid.\cite{Bergara} In Section II we present, within first-order
perturbation [or, equivalently, linear-response] theory, general
expressions for the energy-loss probability of charged particles moving along a
definite trajectory in inhomogeneous media, and focus on the case of a bounded
three-dimensional electron gas that is translationally invariant in the plane of
the surface. In Section III, we report the results of our full self-consistent
calculations of energy-loss spectra of charged particles moving parallel with
the surface, which are found to satisfy sum-rules for particle-number
conservation. Special emphasis is placed on the various contributions from
collective and electron-hole excitations to the energy-loss probability. The
effect of the electronic selvage at a metal surface on the energy-loss spectra
is also discussed, by comparing our full self-consistent calculations with those
obtained for electron densities that drop abruptly to zero at the surface. In
Section IV our conclusions are presented.

Unless otherwise is stated, we use atomic units throughout, i.e.,
$e^2=\hbar=m_e=1$.
 
\section{Theory}

We consider a recoiless particle of charge $Z_1$ moving in an arbitrary
inhomogeneous electron system at a given impact vector ${\bf b}$ with
non-relativistic velocity ${\bf v}$, for which retardation effects and
radiation losses can be neglected.\cite{note1} Within first-order
perturbation theory, the probability for the probe particle to transfer momentum
${\bf q}$ to the medium is given by the following expression:\cite{Pitarke3}
\begin{eqnarray}\label{eq51p} 
P_{\bf q}=&&-{4\pi\over LA}\,Z_1^2\int_0^\infty d\omega
\int{d{\bf q}'\over(2\pi)^3}\,{\rm e}^{i{\bf
b}\cdot({\bf q}+{\bf q}')}\cr\cr
&&\times {\rm Im}W({\bf q},{\bf q}';\omega)\,\delta(\omega-{\bf q}\cdot{\bf
v})\,\delta(\omega+{\bf q}'\cdot{\bf v}),
\end{eqnarray}
where $L$ and $A$ represent the normalization length and area, respectively,
and $W({\bf q},{\bf q}';\omega)$ is the screened interaction
\begin{equation}
W({\bf q},{\bf q}';\omega)=\int d{\bf r}\int d{\bf r}'\,{\rm e}^{-i({\bf
q}\cdot{\bf r}+{\bf q}'\cdot{\bf r}')}\,W({\bf r},{\bf r}';\omega),
\end{equation}
with
\begin{eqnarray}\label{screened}
W({\bf r},{\bf r}';&&\omega)=v({\bf r},{\bf r}')
+\int d{\bf r}_1\int d{\bf r}_2\cr\cr
&&\times v({\bf r},{\bf r}_1)\,\chi({\bf r}_1,{\bf
r}_2,\omega)\,v({\bf r}_2,{\bf r}').
\end{eqnarray}
Here, $v({\bf r},{\bf r}')$ represents the bare Coulomb interaction and
$\chi({\bf r},{\bf r}',\omega)$ is the so-called density-response function of
the medium.\cite{Pines2}  

Within a self-energy formalism, the decay rate of the probe particle is
obtained from the knowledge of the imaginary part of the self-energy. In the GW
approximation, and replacing the probe-particle Green function by that of a
non-interacting recoiless particle, one finds:\cite{review}
\begin{eqnarray}\label{auto4}
\tau^{-1}=&&-2\,Z_{1}^{2}\sum_f\int d{\bf r}\int d{\bf r'}
\,\phi_i^\ast({\bf r})\,\phi_{f}^\ast({\bf r'})\cr\cr
&&\times {\rm Im}W({\bf r},{\bf r'},E_i-E_f)\,\phi_{i}({\bf r'})
\,\phi_{f}({\bf r}),
\end{eqnarray}
where $\phi_i({\bf r})$ represents the probe-particle initial state of energy
$E_i$, and the sum is extended over a complete set of final states $\phi_f({\bf
r})$ of energy $E_f$. Describing the probe-particle
initial and final states by plane waves in the direction of motion and
a Dirac $\delta$ function in the transverse direction, i.e.,
\begin{equation}\label{auto8}
\phi({\bf r})={1\over\sqrt A}\,e^{i{\bf v}\cdot{\bf
r}}\,\sqrt{\delta({\bf r}_\perp-{\bf b})},
\end{equation}
where ${\bf r}_\perp$ represents the position vector perpendicular to the
projectile velocity, one finds  
\begin{equation}\label{tau}
\tau^{-1}={1\over T}\sum_{\bf q}\,P_{\bf q},
\end{equation}
$T$ being the normalization time and $P_{\bf q}$ the probability for the
probe particle to transfer momentum ${\bf q}$ to the medium, as
obtained from Eq. (\ref{eq51p}). 

Alternatively, one may consider the energy that the probe particle looses per
unit time due to electronic excitations in the medium. This can
be written as\cite{Flores1}
\begin{equation}\label{flores}
-{dE\over dt}=-\int d{\bf r}\,\rho^{ext}({\bf r},t)\,{\partial V^{ind}({\bf
r},t)\over\partial t},
\end{equation}
where $\rho^{ext}({\bf r},t)$ represents the probe-particle charge density
\begin{equation}\label{rho}
\rho^{ext}({\bf r},t)=Z_1\,\delta({\bf r}-{\bf b}-{\bf v}t),
\end{equation}
and $V^{ind}({\bf r},t)$ is the induced potential. To first order in
$\rho^{ext}({\bf r},t)$, i.e., within linear-response theory, one finds
\begin{eqnarray}\label{uve}
V^{ind}({\bf r},t)=\int&&d{\bf
r}'\int_{-\infty}^{+\infty} dt'\int_{-\infty}^{+\infty}{d\omega\over 2\pi}\,{\rm
e}^{-i\omega(t-t')}\cr\cr &&\times\left[W({\bf r},{\bf r}',\omega)-v({\bf
r},{\bf r}')\right]\,\rho^{ext}({\bf r}',t').
\end{eqnarray}
Introduction of Eqs. (\ref{rho}) and (\ref{uve}) into Eq.
(\ref{flores}) yields the total energy lost by the particle
\begin{equation}\label{loss}
-\Delta E=\int_{-\infty}^{+\infty} dt\left(-{dE\over dt}\right)=
\sum_{\bf q}\left({\bf q}\cdot{\bf v}\right)P_{\bf q},
\end{equation}
where $P_{\bf q}$ is, as in Eq. (\ref{tau}), the probability of Eq. (\ref{eq51p})
for the probe particle to transfer momentum ${\bf q}$ to the medium, and ${\bf
q}\cdot{\bf v}$ represents the corresponding energy transfer. 

The results in Eqs. (\ref{eq51p}), (\ref{tau}), and (\ref{loss}) are general
expressions for the case of a classical trajectory in an arbitrary inhomogeneous
electron system characterized by the density-response function $\chi({\bf r},{\bf
r}';\omega)$. In particular, in the case of a bounded three-dimensional electron
gas that is translationally invariant in two directions, which we take
to be normal to the $z$ axis, the energy loss of Eq. (\ref{loss}) may be expressed
in terms of the two-dimensional Fourier transform of the screened interaction,
as follows
\begin{eqnarray}\label{general}
-\Delta E=&&-{Z_1^2\over\pi}\int {d{\bf
q}_\parallel\over(2\pi)^2}\int_{-\infty}^{+\infty} dt\int_{-\infty}^{+\infty} dt'
\int_0^\infty d\omega\,\omega\cr\cr
&&\times{\rm e}^{-i(\omega-{\bf q}_\parallel\cdot{\bf v}_\parallel)(t-t')}\,{\rm
Im}W[z(t),z(t');{\bf q}_\parallel,\omega],
\end{eqnarray}
where ${\bf q}_\parallel$ and ${\bf v}_\parallel$ are the momentum transfer 
and the velocity in
the plane of the surface and $z(t)$ represents the position of the projectile
relative to the surface. Eq. (\ref{general}) gives the energy that a charged
particle moving with constant velocity along an arbitrary trajectory looses due
to electronic excitations in an electron system that is translationally
invariant in two directions, as occurs in the case of a simple metal surface
modeled by jellium.

\subsection{Parallel trajectory}

In the glancing incidence geometry ions penetrate into the material, they skim
the outermost layer of the solid, and are then repelled by a repulsive, screened
Coulomb potential, as discussed by Gemmell.\cite{Gemmell} Through use of the
appropriate effective potentials the ion trajectory $z(t)$ can be calculated and
the energy loss is then obtained from Eq. (\ref{general}). Here we restrict our
attention to the case of charged particles moving with constant velocity ${\bf
v}$ along a definite trajectory at a fixed distance $z$ from a jellium surface,
as approximately occurs under extreme grazing-incidence conditions. Eq.
(\ref{general}) then  yields
\begin{equation}
-\Delta E=L\,\left(-{dE\over dx}\right),
\end{equation}
where $\left(-dE/dx\right)$ is the energy loss per unit path length of the
projectile, i.e., the so-called stopping power of the electron system,
\begin{eqnarray}\label{stop}
-{dE\over dx}=&&-{2\over v}\,Z_1^2\int{d{\bf
q}_\parallel\over(2\pi)^2}\int_0^\infty d\omega\,\omega\cr\cr
&&\times{\rm Im}W(z,z;{\bf q}_\parallel,\omega)\,\delta(\omega-{\bf
q}_\parallel\cdot{\bf v}).
\end{eqnarray}

Eq. (\ref{stop}) can be expressed in terms of $P(q_\parallel,\omega)$, 
which represents the probability per unit time, unit wave number and unit
frequency for the probe particle to transfer  momentum
$q_\parallel$  and energy $\omega$  to the medium:
\begin{equation}\label{one}
-{dE\over dx}={1\over v}\int_0^\infty\ dq_\parallel\int_0^{q_\parallel v}
d\omega\,\omega\,P(q_\parallel,\omega), 
\end{equation}
where 
\begin{equation}\label{pqw}
 P(q_\parallel,\omega)=-{Z_1^2\over\pi^2 v}
\,{\rm Im}W(z,z;q_\parallel,\omega)\,
\frac{q_\parallel}{\sqrt{q_\parallel^{2}-(\omega/v)^{2}}}.
\end{equation}

Alternatively, the stopping power of the system is often described
by means of  $P(\omega)$, the total probability of exchanging energy $\omega$
with the medium:
\begin{equation}\label{two}
-{dE\over dx}={1\over v}\int_0^\infty d\omega\,\omega\,P(\omega),
\end{equation}
where
\begin{equation}\label{ptwo}
P(\omega)=-{Z_1^2\over\pi^2 v}\int_0^\infty dq_x\,
{\rm Im}W(z,z;q_\parallel,\omega),
\end{equation}
with $q_\parallel=\sqrt{q_x^2+(\omega/v)^2}$.

The main ingredient in the evaluation of Eqs. (\ref{pqw}) and (\ref{ptwo})
is the screened interaction $W(z,z';q_\parallel,\omega)$. From Eq.
(\ref{screened}), one easily finds
\begin{eqnarray}\label{screened2}
W(z,z';&&q_\parallel,\omega)=v(z,z',q_\parallel)
+\int dz_1\int dz_2\cr\cr
&&\times
v(z,z_1;q_\parallel)\,\chi(z_1,z_2;q_\parallel,\omega)\,v(z_2,z';q_\parallel),
\end{eqnarray}
where $v(z,z';q_\parallel)$ and $\chi(z,z';q_\parallel,\omega)$ are
two-dimensional Fourier transforms of the bare Coulomb interaction and the
density-response function, respectively.

In particular, for $z$ and $z'$ coordinates that are well inside the solid,
there is translational invariance in the direction normal to the surface and
$W(z,z';q_\parallel,\omega)$ can then be easily obtained as follows
\begin{equation}\label{bulk0}
W(z,z';q_\parallel,\omega)=\int_{0}^{\infty}dq_z\, 
e^{iq_z(z-z')}\,v(q)\,\epsilon^{-1}(q,\omega),
\end{equation}
where $q=\sqrt{q_\parallel^2+q_z^2}$ and
$\epsilon^{-1}(q,\omega)$ represents the inverse dielectric function of
a uniform electron gas,
\begin{equation}\label{inverse}
\epsilon^{-1}(q,\omega)=1+v(q)\,\chi(q,\omega),
\end{equation}
$v(q)$ and $\chi({\bf q},\omega)$ being three-dimensional Fourier transforms of
the bare Coulomb interaction and the density-response function, respectively. 

For $z$ and $z'$ coordinates that are far from the surface into the vacuum, where
the electron density vanishes, Eq. (\ref{screened2}) yields
\begin{equation}\label{winf}
W(z,z';q_\parallel,\omega)=v(z,z';q_\parallel)-{2\pi\over
q_\parallel}\,e^{-q_\parallel(z+z')}\,g(q_\parallel,\omega),
\end{equation}
where $g(q_\parallel,\omega)$ is the so-called surface-response function
\begin{equation}\label{gqw}
g(q_\parallel,\omega)=-{2\pi\over
q_\parallel}\int dz_1\int dz_2\,e^{q_\parallel(z_1+z_2)}
\,\chi(z_1,z_2;q_\parallel,\omega).
\end{equation}
The energy-loss function ${\rm Im}g(q_\parallel,\omega)$ satisfies the $f$
sum-rule\cite{Liebsch}
\begin{equation}\label{sumrule}
\int_0^\infty d\omega\,\omega\,{\rm
Im}g(q_\parallel,\omega)=2\,\pi^2\,q_\parallel\int_{-\infty}^{\infty} dz\,{\rm
e}^{2q_\parallel z}\,n(z),
\end{equation}
which applies to the case of a bounded three-dimensional electron gas whose
exact density in the ground state is $n(z)$. For $z$ coordinates that are well
inside the solid the electron density takes a constant value $\bar n$, and for
$z$ coordinates that are far from the surface into the vacuum the electron
density vanishes.

In the long-wavelength limit ($q_\parallel\to 0$),
\begin{equation}\label{long}
{\rm Im}g(q_\parallel,\omega)\to{\pi\over 2}\,\omega_s\,\delta(\omega-\omega_s),
\end{equation}
where $\omega_s=\omega_p/\sqrt{2}$ and $\omega_p={(4\pi\bar n)}^{1/2}$ is the
classical plasma frequency of a uniform electron gas of density $\bar n$.
Hence, in the $q_\parallel\to 0$ limit the energy loss is dominated by the
excitation of surface plasmons of energy $\omega_s$, as predicted by
Ritchie.\cite{Ritchie1}

\subsection{The density-response function}
\label{density}

We consider a jellium slab of thickness $a$ normal to the $z$ axis, consisting of a
fixed uniform positive background of density
\begin{equation}
n_+(z)=\cases{\bar n,&$-a\leq z\leq 0$\cr\cr 0,& elsewhere,}
\end{equation}
plus a neutralizing cloud of interacting electrons of density $n(z)$.
The positive-background charge density $\bar n$ is often expressed in terms of
the Wigner radius $r_s$, as $1/\bar n=(4\pi/3)r_s^3$.

Time-dependent density-functional theory (TDDFT) shows that the {\it exact}
density-response function of the electron system satisfies the integral
equation\cite{tddft}
\begin{eqnarray}\label{chi}
\chi&&(z,z';q_\parallel,\omega)=
\chi^0(z,z';q_\parallel,\omega)+\int dz_1\int{\rm
d}z_2\cr\cr
&&\times\chi^0(z,z';q_\parallel,\omega)\left[v(z_1,z_2;q_\parallel)
+f_{xc}(z_1,z_2;q_\parallel,\omega)\right]\cr\cr
&&\times\chi(z_2,z';q_\parallel,\omega),
\end{eqnarray}
where $\chi^0(z,z';q_\parallel,\omega)$ is the density-response function
of non-interacting Kohn-Sham electrons
\begin{eqnarray}\label{chi0}
\chi^0(&&z,z';q_\parallel,\omega)=2\,\sum_{i,j}\phi_i(z)\phi_{j}^*(z)
\phi_{j}(z')\phi_i^*(z')\cr\cr
&&\times\int{{\rm d}{\bf
k}_\parallel\over(2\pi)^2}\,{\Theta(E_F-E_i)-\Theta(E_F-E_{j})\over
E_i-E_{j}+(\omega+i\eta)},
\end{eqnarray}
and the kernel $f_{xc}(z,z';q_\parallel,\omega)$ accounts for
exchange-correlation (xc) effects beyond a time-dependent Hartree
approximation. In Eq. (\ref{chi0}), $\Theta(x)$ is the Heaviside step function,
$\eta$ is a positive infinitesimal, the energies $E_i$ and $E_j$ are
\begin{equation}\label{4}
E_i=\varepsilon_i+{{\bf k}_\parallel^2\over 2}
\end{equation}
and
\begin{equation}\label{4p}
E_{j}=\varepsilon_{j}+{({\bf k}_\parallel+{\bf q}_\parallel)^2\over 2},
\end{equation}
and the wave functions $\phi_i(z)$ and energies $\varepsilon_i$, which describe
motion normal to the surface, are the eigenfunctions and eigenvalues of
the one-dimensional Kohn-Sham Hamiltonian
\begin{equation}\label{effective}
H=-{1\over 2}\,{d^2\over z^2}+\varphi(z)+v_{xc}(z),
\end{equation}
$\varphi(z)$ being the electrostatic potential and $v_{xc}(z)$ representing
the so-called xc potential of density-functional theory (DFT).\cite{kohn65}

Within this scheme, the simplest possible approximation is to neglect xc
altogether and set the xc potential $v_{xc}(z)$ and the xc kernel
$f_{xc}(z,z';q_\parallel,\omega)$ equal to zero. In this case, the
one-dimensional single-particle wave functions and energies are the
self-consistent eigenfunctions and eigenvalues of the one-electron Hartree
Hamiltonian. The calculation of the density-response function is further
simplified if the self-consistent electrostatic potential entering Eq.
(\ref{effective}) is replaced by
\begin{equation}
\varphi(z)=\cases{0,& $-a-z_0\leq z \leq z_0$\cr\cr
\infty,& elsewhere,}
\end{equation}
where $z_0$ is chosen so as to ensure charge neutrality. This is the so-called
infinite-barrier model (IBM).\cite{Bardeen} Within this model,
the one-electron wave functions are simply sines, and charge neutrality is easily
found to yield \begin{equation}
z_0=(3/16)\,\lambda_F\left[1+O\left(\lambda_F/a\right)\right].
\end{equation}

Exchange-correlation effects are usually introduced within the local-density
approximation (LDA) of DFT, by replacing the xc potential at $z$
by that of a uniform electron gas with the local density $n(z)$. The xc kernel
entering Eq. (\ref{chi}) is then set either equal to zero [this is the
random-phase approximation (RPA)\cite{note2}] or equal to the static
($\omega=0$) xc kernel
\begin{equation}\label{kernel}
f_{xc}^{ALDA}(z,z';q_\parallel,\omega)=\left[{dv_{xc}(n)\over
dn}\right]_{n=n(z)}\,\delta(z-z').
\end{equation}
This is the so-called adiabatic local-density approximation (ALDA).\cite{tdlda}

To compute the interacting density-response function
$\chi(z,z';q_\parallel,\omega)$, we follow the method described in
Ref.\onlinecite{Eguiluz}. We first assume that
$n(z)$ vanishes at a distance
$z_0$ from either jellium edge,\cite{note3} and expand the wave functions
$\phi_i(z)$ in a Fourier sine series. We then introduce a double-cosine Fourier
representation for the density-response function, and find explicit expressions
for the screened interaction and the surface-response function in terms of the
Fourier coefficients of the density-response function (see Appendix A).

Great care was exercised to ensure that our slab calculations are a faithful
representation of the screened interaction and the energy-loss probability in a
semi-infinite medium. This issue is important, in view of the significant
quantum-size effects (QSE)\cite{Schulte} originated in the quantization of the
energy levels normal to the surface: as the slab-thickness $a$ increases new
subbands for the $z$ motion become occupied, thereby leading to oscillatory
functions of $a$ (the amplitude of these oscillations decays approximately
linearly with $a$, and their period equals $\lambda_F/2$,
$\lambda_F=2\pi/(3\pi^2\bar n)^{1/3}$ being the Fermi wavelength). For each
quantity
$\alpha$ under study we considered three different values of $a$. One such value
is the threshold width $a_n$ for which the $n$th subband for the $z$ motion is
first occupied. The other two values are $a_n^-=a_n-\lambda_F/4$ and
$a_n^+=a_n+\lambda_F/4$, and the infinite-width limit is then extrapolated with
the use of the following relation\cite{qsefor,qsefor2}
\begin{equation}
\alpha={\alpha(a_n^-)+\alpha(a_n)+\alpha(a_n^+)\over 3}. 
\end{equation} 
Following this procedure to calculate the 
surface-response function $g(q_\parallel,\omega)$, we have been able
to prove that the sum-rule of Eq. (\ref{sumrule}) is satisfied for all values
of $q_\parallel$ under consideration. The results presented below correspond to
slabs with $n=12$, for which $a\approx 5-6\lambda_F$.

\subsection{Simplified models}

For comparison, we also consider various simplified models for the screened
interaction $W(z,z';q_\parallel,\omega)$ of a semi-infinite free-electron gas,
which are all derived for electron densities that drop abruptly to zero at the
surface. These are: (a) a classical model consisting of a semi-infinite medium
of local dielectric function $\epsilon(\omega)$, (b) semiclassical and quantized
hydrodynamic (HD) models,\cite{Wilems,Barton} and (c) a classical infinite
barrier (CIB)  or  specular-reflexion (SR) model,\cite{Reuter,Marusak} which has
the   virtue of incorporating dispersion effects by expressing the screened
interaction in terms of the bulk dielectric function $\epsilon(q,\omega)$.
Within these models and for $z$ and $z'$ coordinates that are outside the
surface ($z,z'>0$) the screened interaction is obtained through Eq.
(\ref{winf}), from the knowledge of approximate expressions for the
surface-response function $g(q_\parallel,\omega)$.

\subsubsection{Classical model}

Within this approach, the screened interaction is derived by 
imposing the ordinary boundary conditions of continuity of the potential and
the normal component of the displacement vector at the surface ($z=0$). For
$z,z'>0$, one then easily finds Eq. (\ref{winf}) with
\begin{equation}\label{clas}
g(\omega)={\epsilon(\omega)-1\over\epsilon(\omega)+1}.
\end{equation}

For a free-electron gas, the long-wavelength ($q\to 0$) dielectric function is
\beq\label{feg}
\epsilon(\omega)=1-\frac{\omega_{p}^{2}}{\omega(\omega+i\eta)},
\eeq
and introduction of Eq. (\ref{feg}) into Eq. (\ref{clas}) yields the
long-wavelength limit of Eq. (\ref{long}). Introducing this limit into either Eq.
(\ref{pqw}) or Eq. (\ref{ptwo}), one easily reproduces the classical expression
of Echenique and Pendry\cite{Pendry} for the stopping power of a jellium surface,
\beq\label{stopcl}
-{dE\over dx}=Z_1^2\,{\omega_s^2\over v^2}\,
{\rm K}_0(2\,\omega_s\,z/v),
\eeq
where ${\rm K}_0$ is the zero-order modified Bessel function.
For large values of $z$ ($z>>v/\omega_s$), Eq. (\ref{stopcl}) reduces to
\beq\label{stopcllargez}   
-{dE\over dx}=Z_1^2\,{\omega_s\over 2\,v}\,
\sqrt{\pi\,\omega_s/z\,v}\,\,e^{-2\,\omega_s\,z/v}.
\eeq 

\subsubsection{Hydrodynamic models}

In a HD model, the collective motion of electrons in an arbitrary inhomogeneous
system is expressed in terms of the deviations from the equilibrium density. In
a semiclassical approach, one writes and linearizes the basic hydrodynamic
equations, i.e., the continuity and the Bernuilli equation, and for a
semi-infinite system finds
\begin{equation}\label{hdc}
g(q_\parallel,\omega)={\omega_p^2\over
2\beta^2\Lambda_{q_\parallel}(\Lambda_{q_\parallel}+q_\parallel)-\omega_p^2},
\end{equation}
where
\begin{equation}
\Lambda_{q_\parallel}={1\over\beta}\sqrt{\omega_p^2+\beta^2q_\parallel^2-
\omega(\omega+i\eta)}
\end{equation}
and $\beta$ represents the speed of propagation of hydrodynamic disturbances in
the electron system.\cite{note4}

Within a quantized hydrodynamic model, one first linearizes the Hamiltonian of
the hydrodynamic system with respect to the induced electron density, and then
quantizes this Hamiltonian on the basis of the normal modes of oscillation,
which are referred after quantization as bulk and surface plasmons. Hence,
within this approach one can distinguish the separate contributions to the
energy-loss function ${\rm Im}g(q_\parallel,\omega)$ coming from the excitation
of either bulk or surface plasmons:\cite{Aitor}
\begin{eqnarray}\label{hdp}
{\rm Im}g^B(q_\parallel,\omega)=&&{1\over 2}\,q_\parallel\,\int_0^\infty dq_z
\,\delta(\omega-\omega_q^B)\cr\cr
&&\times{(\omega_p^2/\omega_q^B)\,q_z^2\over
q_z^4+q_z^2(q_\parallel^2+\omega_p^2/\beta^2)+\omega_p^4/(4\beta^4)}
\end{eqnarray}
and
\begin{equation}\label{hds}
{\rm Im}g^S(q_\parallel,\omega)={\pi\over 2}\,{\gamma_{q_\parallel}\over
q_\parallel
+2\gamma_{q_\parallel}}\,{\omega_p^2\over\omega_{q_\parallel}^S}\,
\delta(\omega-\omega_{q_\parallel}^S),
\end{equation}
respectively. Here, $\omega_q^B$ and $\omega_{q_\parallel}^S$ represent the
dispersion of bulk and surface plasmons,
\begin{equation}
\left(\omega_q^B\right)^2=\omega_p^2+\beta^2\,q^2
\end{equation}
and
\begin{equation}
\left(\omega_{q_\parallel}^S\right)^2={1\over 2}\left[\omega_p^2+
\beta^2\,q_\parallel^2+\beta\,q_\parallel\,
\sqrt{2\omega_p^2+\beta^2\,q_\parallel^2}\right].
\end{equation}
As in Eq. (\ref{bulk0}) $q=\sqrt{q_\parallel^2+q_z^2}$, and 
\begin{equation}
\gamma_{q_\parallel}={1\over 2\beta}\left(-\beta
q_\parallel+\sqrt{2\omega_p^2+\beta^2q_\parallel^2}\right).
\end{equation}

For the separate contributions  to the sum-rule  of  Eq. (\ref{sumrule})
coming from bulk and surface plasmons, integration of Eqs. (\ref{hdp}) and
(\ref{hds}) yields
\begin{equation}\label{sumruleb}
\int_0^\infty d\omega\,\omega\,{\rm
Im}g^B(q_\parallel,\omega)={\pi\over 4}\,{q_\parallel\over
q_\parallel
+2\gamma_{q_\parallel}}\,\omega_p^2
\end{equation}
and
\begin{equation}\label{sumrules}
\int_0^\infty d\omega\,\omega\,{\rm
Im}g^S(q_\parallel,\omega)={\pi\over 4}\,{2\gamma_{q_\parallel}\over
q_\parallel
+2\gamma_{q_\parallel}}\,\omega_p^2,
\end{equation}
respectively. It is then straightforward to show that for a semi-infinite
system with a uniform electron density $\bar n$ that drops abruptly to zero at
the surface the sum of bulk and surface contributions to the energy-loss
function ${\rm Im}g(q_\parallel,\omega)$ satisfies Eq. (\ref{sumrule}).

In the limit as $q_\parallel\to 0$ the bulk contribution to the energy-loss
function vanishes, and both the imaginary part of Eq. (\ref{hdc}), on the one
hand, and Eq. (\ref{hds}), on the other hand, yield the long-wavelength limit of
Eq. (\ref{long}).

\subsubsection{Specular reflexion model}

Either by neglecting, within the IBM, the interference between incident and
scattered electrons (CIBM),\cite{Reuter} or by simply assuming that electrons are
specularly reflected at the surface (SRM)\cite{Marusak}, one finds
\beq\label{gsrm}
g(q_\parallel,\omega)=\frac{1-\epsilon_{s}(q_\parallel,\omega)}{1+\epsilon_
{s}(q_\parallel,\omega)},
\eeq
where
\beq\label{srm}
\epsilon_{s}(q_\parallel,\omega)=\frac{q_\parallel}{\pi}\int_{-\infty}^{\infty}
\frac{dq_{z}}{q^2}\epsilon^{-1}(q,\omega),
\eeq
with $q=\sqrt{q_\parallel^2+q_z^2}$ and $\epsilon^{-1}(q,\omega)$ being the
inverse bulk dielectric function of Eq. (\ref{inverse}).

If dispersion effects are neglected altogether, thereby replacing the
momentum-dependent dielectric function
$\epsilon(q,\omega)$ entering Eq. (\ref{srm}) by a local dielectric
function $\epsilon(\omega)$, Eq. (\ref{gsrm}) yields the classical prediction
[Eq. (\ref{clas})]. Alternatively, if dispersion effects are incorporated in
an approximated manner through the hydrodynamic dielectric function of a
uniform electron gas,
\begin{equation}
\epsilon(q,\omega)=1+{\omega_p^2\over\beta^2\,q^2-\omega(\omega+i\eta)},
\end{equation}
Eq. (\ref{gsrm}) is easily found to yield the hydrodynamic surface-response
function [Eq. (\ref{hdc})].

\section{Results and discussion}

We choose the bulk charge density $\bar n$ to be equal to the average electron
density of valence electrons in aluminum metal ($r_s=2.07$),
for which the Fermi momentum [$q_F=(3\pi^2\bar n)^{1/3}$] and bulk plasma
frequency [$\omega_p$] are $q_{F}=0.927\,a_0^{-1}$ ($a_0$ is the Bohr radius,
$a_0=0.529\,{\rm\AA}$) and $\omega_p=15.8$ eV, respectively. We set $Z_1=\pm 1$
and our results can then be used for arbitrary values of $Z_{1}$, as the
energy-loss probability is, within linear-response theory, proportional to
$Z_1^2$.                         

In this section, we first show results for the energy-loss function
${\rm Im}W(z,z;q_\parallel,\omega)$ entering Eqs. (\ref{pqw}) and
(\ref{ptwo}). Fig. 1 shows ${\rm Im}W(z,z;q_\parallel,\omega)$, as a function
of $\omega$, with $q_\parallel=0.4\,q_F$ (for this small value of $q_\parallel$
both bulk and surface plasmons are well defined excitations) and $z$ coordinates
that are either well inside the solid [$z\le-\lambda_F$] (Fig. 1a) or far from
the surface into the vacuum [$z\ge\lambda_F$] (Fig. 1b). We have carried out
self-consistent RPA and ALDA slab calculations for this quantity, as described
in Sec. II.B, and have found the expected result that for these values of $z$
they coincide with those obtained from Eqs. (\ref{bulk0}) [$z\le-\lambda_F$] and
(\ref{winf}) [$z\ge\lambda_F$]. Hence, the energy-loss function ${\rm
Im}W(z,z;q_\parallel,\omega)$ represented in Fig. 1 either does not depend on
$z$ (Fig. 1a) or depends on $z$ through an overall factor of ${\rm
e}^{-2q_\parallel z}$ (Fig. 1b).

In Fig. 1a, bulk-plasmon (dashed line) and electron-hole-pair (dotted line)
contributions to the RPA energy-loss function 
${\rm Im}W(z,z;q_\parallel,\omega)$
are shown separately, as obtained from Eq. (\ref{bulk0}), together with the total
energy-loss function represented by a solid line. One sees that inside the solid
the energy-loss spectrum is dominated, for small values of $q_\parallel$, by a
continuum of bulk-plasmon excitations occurring at energies
$\omega_{q_\parallel}^B<\omega<\omega_{q_c}^B$, where $q_c$ represents the
critical momentum for which the bulk-plasmon dispersion
$\omega_q^B$ enters the electron-hole-pair excitation spectrum. For $r_s=2.07$
and $q_\parallel=0.4\,q_F$, one finds $\omega_{q_\parallel}^B=17.6\,{\rm eV}$
and $\omega_{q_c}^B=23.6\,{\rm eV}$, and bulk plasmons can be excited by
charged particles moving parallel with the surface with speed $v>1.13\,v_0$
($v_0$ is the Bohr velocity, $v_0=2.19\times10^6\,{\rm ms^{-1}}$). 

For $z$ coordinates that are outside the surface it had been generally believed
that the energy loss originates entirely in the excitation of surface plasmons
and electron-hole pairs.\cite{review0} Nevertheless, the continuum of
bulk-plasmon excitations dominating the energy loss of charged particles moving
inside the solid (see Fig. 1a) is still present for $z$ coordinates outside the
surface, as shown in Fig. 1b, although the main contribution to the energy loss
now clearly comes from the excitation of surface plasmons at
$\omega=\omega_{q_\parallel}^S$ [for $r_s=2.07$ and $q_\parallel=0.4\,q_F$, one
finds $\omega_{q_\parallel}^S\sim 16.0\,{\rm eV}$]. The bulk-plasmon
contribution to ${\rm Im}\,W(z,z,q_\parallel,\omega)$, as obtained for
$z=\lambda_F$ within a quantized hydrodynamic model by introducing Eq.
(\ref{hdp}) into Eq. (\ref{winf}) (see also Ref.\onlinecite{Aitor}), is
represented in Fig. 1b by a dotted line. The total SRM energy-loss function, as
obtained with the use of Eq. (\ref{gsrm}) and the RPA dielectric function
$\epsilon(q,\omega)$, is represented in Fig. 1b by a solid line. This curve
shows that at low frequencies the energy-loss spectrum is dominated by the
creation of electron-hole pairs, losses centered around $\omega_{q_\parallel}^S$
are due to the excitation of surface plasmons exhibiting a finite linewidth, and
bulk-plasmon excitations yield energy losses at
$\omega\ge\omega_{q_\parallel}^B$ that nearly coincide with the result one
obtains within the quantized hydrodynamic model. We have also carried out
self-consistent slab calculations of the energy-loss function ${\rm
Im}\,g(q_\parallel,\omega)$, which for small values of $q_\parallel$
($q_\parallel<q_F$) has been found to be close to that obtained within
the SRM. Nevertheless, differences have been found in the surface-plasmon
energy $\omega_{q_\parallel}^S$, which shifts to lower frequencies as
demonstrated earlier;\cite{Tsuei1,Tsuei2} also, the thickness of the slab is
required to be very large for the self-consistent calculations to properly
account for the high-energy spectrum originated in the excitation of bulk
plasmons and electron-hole pairs.

Fig. 2 shows self-consistent calculations of the probability $P(\omega)$ for
a charged particle to exchange energy $\omega$ with the medium, as obtained from
Eq. (\ref{ptwo}). The particle is assumed to move parallel to the surface with
$v=2\,v_0$ and two different $z$ coordinates for which the electronic selvage at
the surface is expected to play an important role: $z=-\lambda_F/2$ (Fig. 2a) and
$z=\lambda_F/2$ (Fig. 2b). Dashed, thick-solid and thin-solid lines give the
result of IBM, RPA and ALDA calculations, and dashed-dotted lines represent
the SRM probability obtained with the RPA dielectric function
$\epsilon(q,\omega)$.

In the interior of the solid (Fig. 2a), the bulk-plasmon energy-loss spectrum
is known to be inhibited by the presence of surface-plasmon losses through
the so-called bregenzung or boundary effect predicted by Ritchie,\cite{Ritchie1}
whose existence is due to the orthogonality of the surface-plasmon and
bulk-plasmon modes. Nevertheless, as the electronic selvage is changed from
zero (SRM) to its actual structure (RPA and ALDA),\cite{note5} the creation of
electron-hole pairs increases, the surface-plasmon peak diminishes, and a broad
bulk-plasmon peak dominates the spectrum, showing that a proper treatment of
the surface density-profile is crucial for the energy dependence of the
energy-loss probability. These results are in qualitative agreement with the
calculations reported in Ref.\onlinecite{Ritchie3}, where the selvage structure
is introduced in an approximate manner within the HD and the SR models.

Bulk-plasmon losses occurring in the vacuum side of the surface (see
Fig. 1b) are not visible in the total energy-loss probability $P(\omega)$
which is dominated by the excitation of surface plasmons and electron-hole pairs,
as shown in Fig. 2b. Also, this figure shows substantial changes in the
energy-loss probability as a realistic description of the surface response is
considered, with an important shift of the surface-plasmon peak towards smaller
energies, in agreement with the experimentally determined surface-plasmon
energies of simple metals.\cite{Tsuei1,Tsuei2} 

Figure 3 shows a surface plot of our full IBM (Fig. 3a), RPA (Fig.
3b) and ALDA (Fig. 3c) calculations of the probability $P(\omega)$ for a charged
particle moving with speed $v=2\,v_0$ parallel to the surface. The plot is shown
as a function of the energy loss $\omega$ and the distance $z$ from the particle
trajectory to the surface. Although the energy-loss probability is found
to be divided into losses centered around the bulk-plasmon energy (in the
interior of the solid) and the surface-plasmon energy (outside the
solid), this separation is not as clear as predicted with the use of
simplified models for the surface response (see, e.g.,
Ref.\onlinecite{Ritchie3}). As $z\to-\infty$ the energy-loss probability
$P(\omega)$ reaches a constant shape centered around the bulk-plasmon energy,
which does not depend on the details of the electronic selvage at the surface and
only depends on whether the xc kernel $f_{xc}(z,z';q_\parallel,\omega)$ is set
equal to zero (IBM and RPA) or not (ALDA). Outside the solid, the energy-loss
probability, which is centered around the surface-plasmon energy, decreases with
the distance $z$ from the surface to the particle trajectory.

Fig. 4a depicts our full IBM (dashed line), RPA (thick-solid line) and ALDA
(thin-solid line) calculations of the stopping power, as obtained from either
Eq. (\ref{one}) or (\ref{two}) as a function of $z$ and with $v=2\,v_0$. In the
interior of the solid, where the electron density is taken to be constant, both
IBM and RPA stopping powers coincide with the well-known RPA stopping power of a
uniform electron gas. Short-range xc effects, included in the ALDA, provoke a
reduction in the screening of electron-electron interactions, thereby increasing
the energy loss. Outside the solid the electronic selvage at the surface plays a
crucial role in the actual behaviour of the stopping power: a slow decrease of
the electron density at the metal surface leads to a larger energy-loss
probability [see also Fig. 2 and Eq. (\ref{sumrule})], and the IBM stopping
power is, therefore, found to be too small. In the SRM the electron density is
assumed to drop abruptly to zero at the surface, which provokes a reduction in
the electron-hole excitation probability, and the stopping power outside the
solid (see  Fig. 4b) is found to be even smaller than in the IBM. Low-energy
excitations involve transitions from occupied electronic states near the Fermi
level, which are sensitive to the actual density profile at the surface, and are
found to play an important role in the energy-loss mechanism of charged
particles moving with $v\le 2\,v_0$. Also plotted in Fig. 4b is the result of
assuming that the stopping power for a charged particle that moves at a distance
$z$ from the surface can be approximated by that of a uniform electron gas with
the local density
$n(z)$.\cite{note6} This often-used local-density approximation also yields an
inaccurate description of the position-dependent stopping power, due to the
intrinsic nature of surface-induced excitations not present within this
approach, and the results presented in Fig. 4 show the need for a
self-consistent description of the surface response if one is to look at the
energy loss of charged particles moving outside a solid surface.

As the velocity increases the energy-loss spectrum of charged particles moving
far from the surface into the vacuum is dominated by long-wavelength
excitations and the stopping power is dictated by the integration of
$\omega\,{\rm Im}g(q_\parallel,\omega)$, which as a result of particle
conservation [see Eq. (\ref{sumrule})] does not depend on the details of the
actual response of the solid. In this limit and with the aid of Eq.
(\ref{sumrule}) one easily finds the classical stopping power of Eq.
(\ref{stopcl}).

The velocity dependence of the stopping power is shown in Fig. 5a, for a
particle moving outside the surface at $z=\lambda_F$. Our full IBM, RPA and
ALDA calculations are represented, as in Fig. 4a, by dashed, thick-solid and
thin-solid lines, respectively, and the SRM stopping power is represented by a
dashed-dotted line, as obtained with the RPA dielectric function
$\epsilon(q,\omega)$. At low velocities the energy-loss spectrum is
dominated by intermediate and short-wavelength excitations, even far from
the surface into the vacuum, and a combination of the actual electronic
selvage at the surface with the intrinsic nature of surface-induced
excitations play an important role in increasing the energy-loss. At high
velocities the energy-loss spectrum is dominated by the surface-plasmon
excitation [see Eq. (\ref{long})] and all  calculations converge with the
classical limit of Eq. (\ref{stopcl}), as shown in  Fig. 5b. As in Fig. 4b, the
local-density approximation is also represented in this figure, showing that
this often-used approximation cannot account for the energy loss originated in
surface-induced excitations, not even at low velocities where the energy loss is
entirely due to the excitation of electron-hole pairs.
                                                                           
\section{summary}

We have reported self-consistent calculations of the energy loss spectra
of charged particles moving parallel to a plane-bounded free-electron gas, in
the framework of linear-response theory.

We have found that the continuum of bulk-plasmon excitations dominating the
energy loss of charged particles moving inside the solid is still present for
particle trajectories outside the surface. Nevertheless, these bulk-plasmon
excitations are found not to be visible in the total energy-loss probability
$P(\omega)$ which outside the solid is clearly dominated by the excitation of
surface plasmons and electron-hole pairs.

As for the effect of the electronic selvage at the surface, we have found that
the so-called bregenzung or boundary effect inside the solid is diminished,
plasmon peaks are broadened, and the surface-plasmon peak is considerably
shifted towards smaller energies. The electronic selvage at the surface has
also been found to increase both the energy-loss probability and the stopping
power for charged particles moving in the vacuum side of the surface.

In the high-velocity limit and for charged particles moving far from the surface
into the vacuum the actual stopping power is found to converge with the
classical limit dictated by Eq. (\ref{stopcl}). However, at low and intermediate
velocities substantial changes in the stopping power have been observed as a
realistic description of the surface response is considered, and we have
concluded that a self-consistent description of the surface response is
necessary if one is to look at the energy loss of charged particles moving
outside a solid surface. Accurate measurements of the energy loss of protons
being reflected from a variety of solid surfaces at grazing incidence have
been reported.\cite{stop1,stop2,stop3} A theoretical description of these
experiments requires that the ion trajectory $z(t)$ be calculated and energy
losses from the excitation of inner-shells be taken into account. Also, in real
experiments band-structure effects might be important and the surface
roughness might lead to additional energy loss due to the so-called
Smith-Purcell effect.\cite{Smith} Work in this direction is now in progress. 

\acknowledgments

The authors would like to thank A. G. Eguiluz and R. H. Ritchie for useful
discussions in connection with this research. Partial support by the University
of the Basque Country, the Basque Unibertsitate eta Ikerketa Saila, and the
Spanish Ministerio de Educaci\'on y Cultura is also acknowledged. 
 
\appendix
\section{}

Here we give explicit expressions for the screened interaction and the
surface-response function in terms of the Fourier coefficients of the
density-response function.

We first introduce the following double-cosine Fourier representation for the
density-response function:
\begin{equation}\label{apend1}
\chi(z,z';q_\parallel,\omega)=\sum_{m=0}^\infty\sum_{n=0}^\infty
\chi_{mn}(q_\parallel,\omega)\,\cos(m\pi\tilde z)\,\cos(n\pi\tilde z'),
\end{equation}
where $d=a+2z_0$ and $\tilde z=(z+z_0+a)/d$.

Introducing Eq. (\ref{apend1}) into Eq. (\ref{screened2}), we obtain the
following expression for the screened interaction:
\begin{equation}
W(z,z';q_\parallel,\omega)=\sum_{m=0}^\infty\sum_{n=0}^\infty
W_{mn}(q_\parallel,\omega)\,\cos(m\pi\tilde z)\,\cos(n\pi\tilde z'),
\end{equation}
where 
\begin{eqnarray}\label{alpha}
W_{mn}(q_\parallel,&&\omega)=v_{mn}(q_\parallel)+{\mu_m\mu_n\over d^2}\cr\cr
&&\times\sum_{m'=0}^\infty\sum_{n'=0}^\infty
v_{mm'}(q_\parallel)\chi_{m'n'}(q_\parallel,\omega)v_{n'n}(q_\parallel),
\end{eqnarray}
\begin{eqnarray}\label{coulomb}
v_{mn}(q_\parallel)=&&{2\pi e^2\over q_\parallel^2+(m\pi/d)^2}
\left[{2d\over\sqrt{\mu_m\mu_n}}\,\delta_{mn}\right.\cr\cr
&&\left.-\left[1+(-1)^{m+n}\right]
{q_\parallel\left[1-(-1)^m{\rm e}^{-q_\parallel
d}\right]\over(q_\parallel^2+(n\pi/d)^2}\right],
\end{eqnarray}
and
\begin{equation}
\mu_m=\cases{1,&for $m=0$\cr\cr
2,&for $m\ge 1$.}
\end{equation}

Similarly, introduction of Eq. (\ref{apend1}) into Eq. (\ref{gqw}) yields:
\beq\label{apend7}
g(q_\parallel,\omega)=-{2\pi d^2\over q_\parallel}\sum_{m=0}^\infty
\sum_{n=0}^\infty{1\over\mu_m\mu_n}\,\alpha_{m}\alpha_{n}
\chi_{mn}(q_\parallel,\omega),
\eeq
where
\begin{equation}
\alpha_m=-{\mu_mq_\parallel\over d}\,{1-e^{q_\parallel d}
\cos(m\pi)\over q_\parallel^2+(m\pi/d)^2}.
\end{equation}

\begin{figure}\label{fig1}
\caption{The energy-loss function, ${\rm Im}W(z,z;q_\parallel,\omega)$, as a
function of $\omega$ with $q_\parallel=0.4\,q_F$ and $z$ coordinates that are
either well inside the solid [$z\le-\lambda_F$] or far from the surface into the
vacuum [$z\ge\lambda_F$]. (a) The solid line represents the total
RPA energy-loss function, as obtained from Eq. (\ref{bulk0}) with
$z\le-\lambda_F$; dashed and dotted lines represent the corresponding
bulk-plasmon and electron-hole-pair contributions, respectively. (b) The solid
line represents the SRM energy-loss function, as obtained for $z=\lambda_F$ with
the use of Eq. (\ref{gsrm}) and the RPA dielectric function $\epsilon(q,\omega)$;
the dotted line represents the bulk-plasmon contribution, as obtained by
introducing Eq. (\ref{hdp}) into Eq. (\ref{winf}).}
\end{figure}

\begin{figure}\label{fig4}
\caption{The energy-loss probability $P(\omega)$, as obtained from Eq.
(\ref{ptwo}) with $v=2\,v_0$ and two different $z$ coordinates: (a)
$z=-\lambda_F/2$ and (b) $z=\lambda_F/2$. Dashed-dotted, dashed,
thick-solid and thin-solid lines represent SRM, IBM, RPA and ALDA
calculations. The SRM probabilities have been obtained with the use of the RPA
dielectric function $\epsilon(q,\omega)$. The damping parameter is
 taken to be 
$\gamma= \omega_{p}/10$.}
\end{figure} 

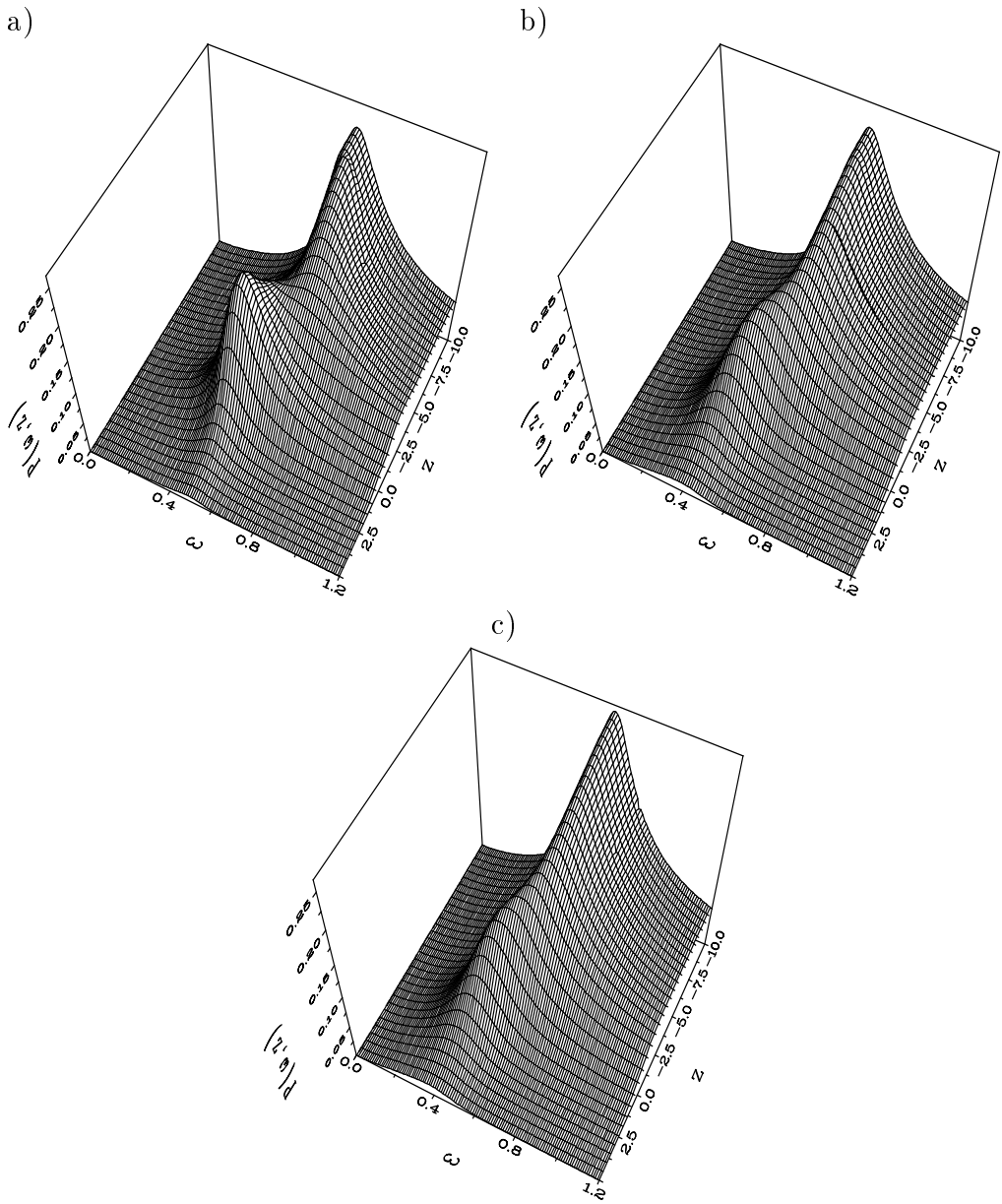
\begin{figure}\label{fig3}
\caption{The energy-loss probability $P(\omega)$, versus the
energy loss $\omega$ and the $z$ coordinate, as obtained for $v=2\,v_0$ within
the (a) IBM, (b) RPA and (c) ALDA. The solid is in the region $z<0$. 
$\gamma= \omega_{p}/10$.}
\end{figure}

\begin{figure}\label{fig5}
\caption{Stopping power, as obtained from either Eq. (\ref{one}) or
(\ref{two}) as a function of $z$ and with $v=2\,v_0$. a) Dashed, thick-solid and
thin-solid lines represent IBM, RPA and ALDA calculations. b)
RPA (thick-solid line), SRM -as obtained with the use of the RPA
dielectric function $\epsilon(q,\omega)$- (dotted line) and local-density (dashed
line) calculations. The solid is in the region $z<0$. $\gamma= \omega_{p}/10$.}
\end{figure}

\begin{figure}\label{fig6}
\caption{Stopping power, as obtained from either Eq. (\ref{one}) or
(\ref{two}) as a function of $v$ and with 
$z=\lambda_F/2$. a) Dashed-dotted, dashed,
thick-solid and thin-solid lines represent SRM -as obtained with the use of the
RPA dielectric function $\epsilon(q,\omega)$-, IBM, RPA and ALDA calculations.
b) IBM (solid line), SRM (dashed-dotted line) and local-density
(dashed line).  The classical prediction of Eq. (\ref{stopcl}) is
represented by a dotted line. 
$\gamma= \omega_{p}/100$.}
\end{figure}


\begin{references}
\bibitem{note1} This approximation is valid for heavy charged particles, e.g.,
ions, and also for swift electrons moving with velocities that are large compared
to those of the electrons in the medium. In the case of electrons, $Z_1=-1$.
\bibitem{note3} In the non-self-consistent IBM, $z_0$ is chosen so as to ensure
charge neutrality. For a self-consistent description of the density-response
function, $z_0$ is chosen sufficiently large for the physical results to be
insensitive to the precise value employed.
\bibitem{note5} Self-consistent RPA and ALDA electron-densities coincide
and both vary very slowly as compared to the IBM. While the IBM
electron density vanishes for $z\ge(3/16)\lambda_F$, self-consistent electron
densities can only be neglected for $z>\lambda_F$ (see, e.g., Fig. 8 of
Ref.\onlinecite{qsefor2}).
\bibitem{note6} The use of this local-density approximation, in which
one replaces $-(dE/dx)(z)$ in either either Eq. (\ref{one}) or (\ref{two}) by the
stopping power of a uniform electron gas of density $n(z)$, should not be
confused with the use of the LDA xc potential in Eqs. (\ref{effective}) and
(\ref{kernel}).
\end{references}
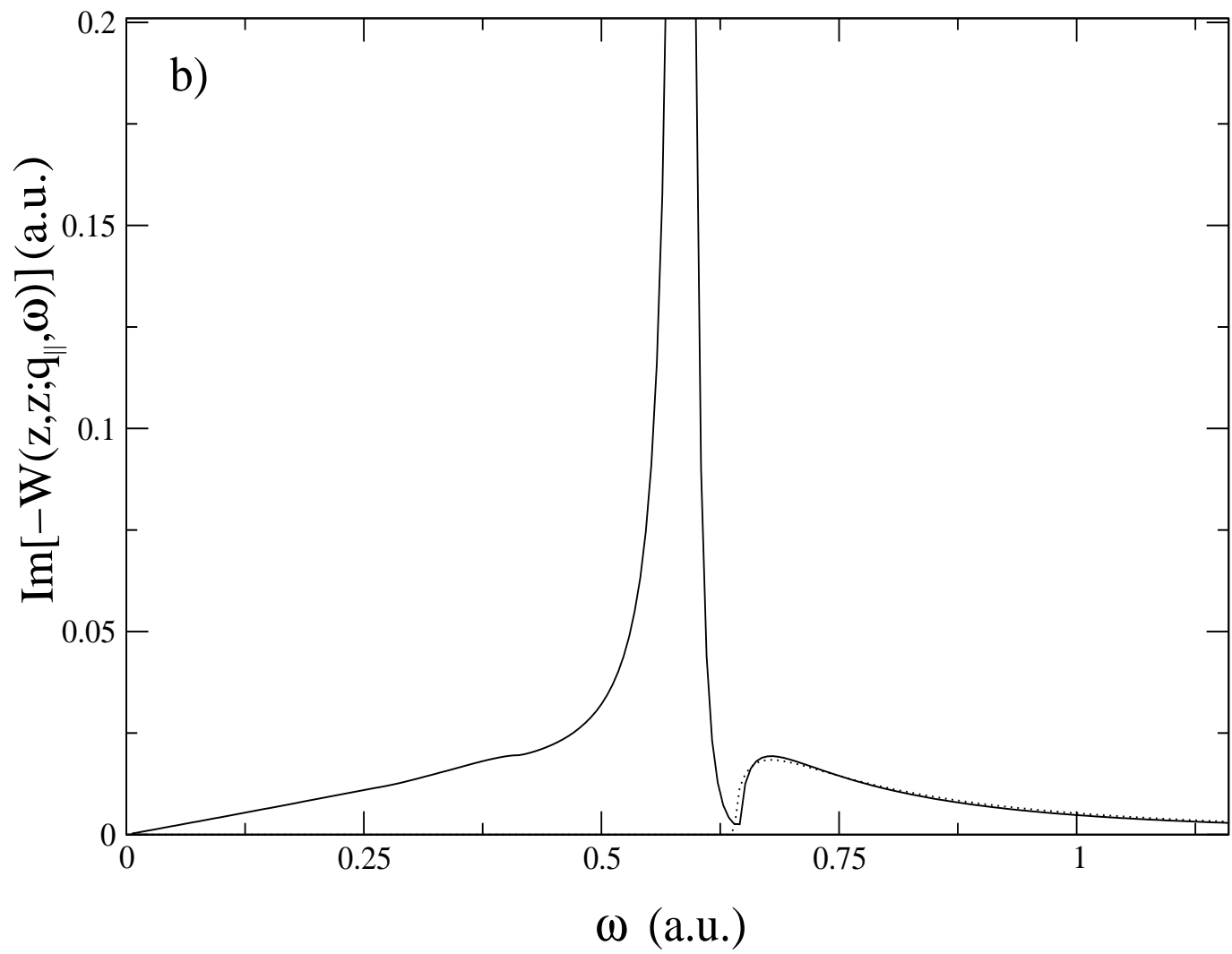
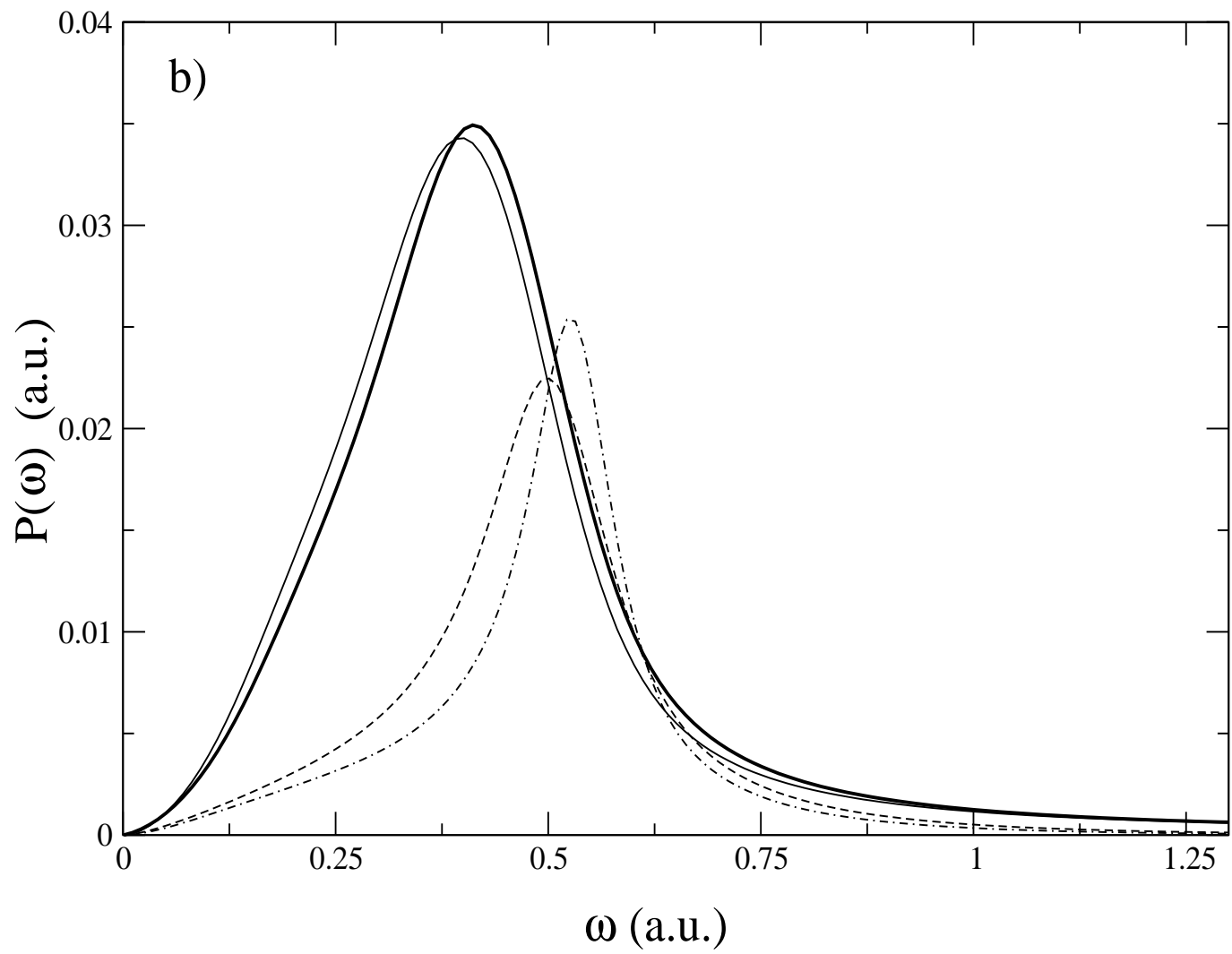
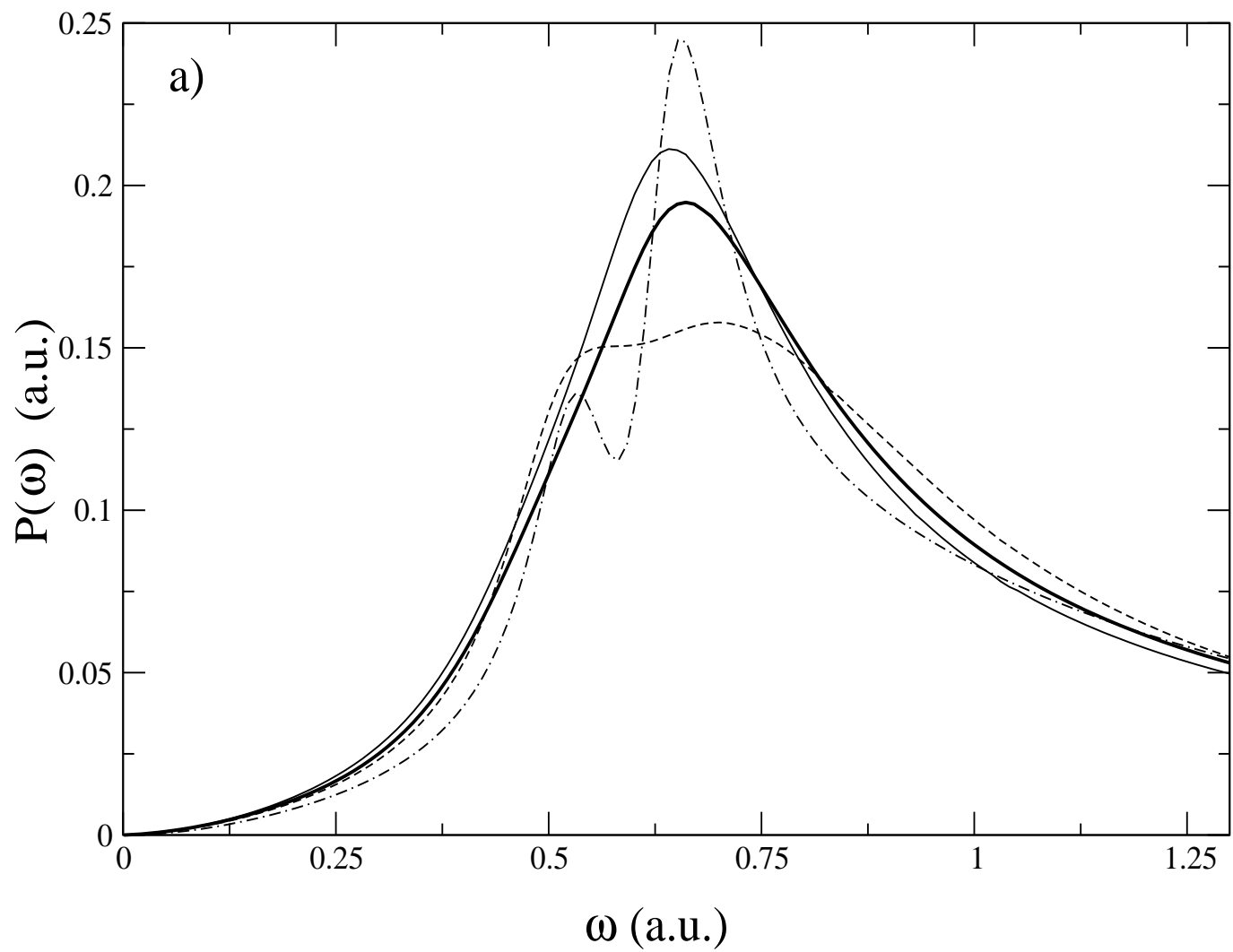
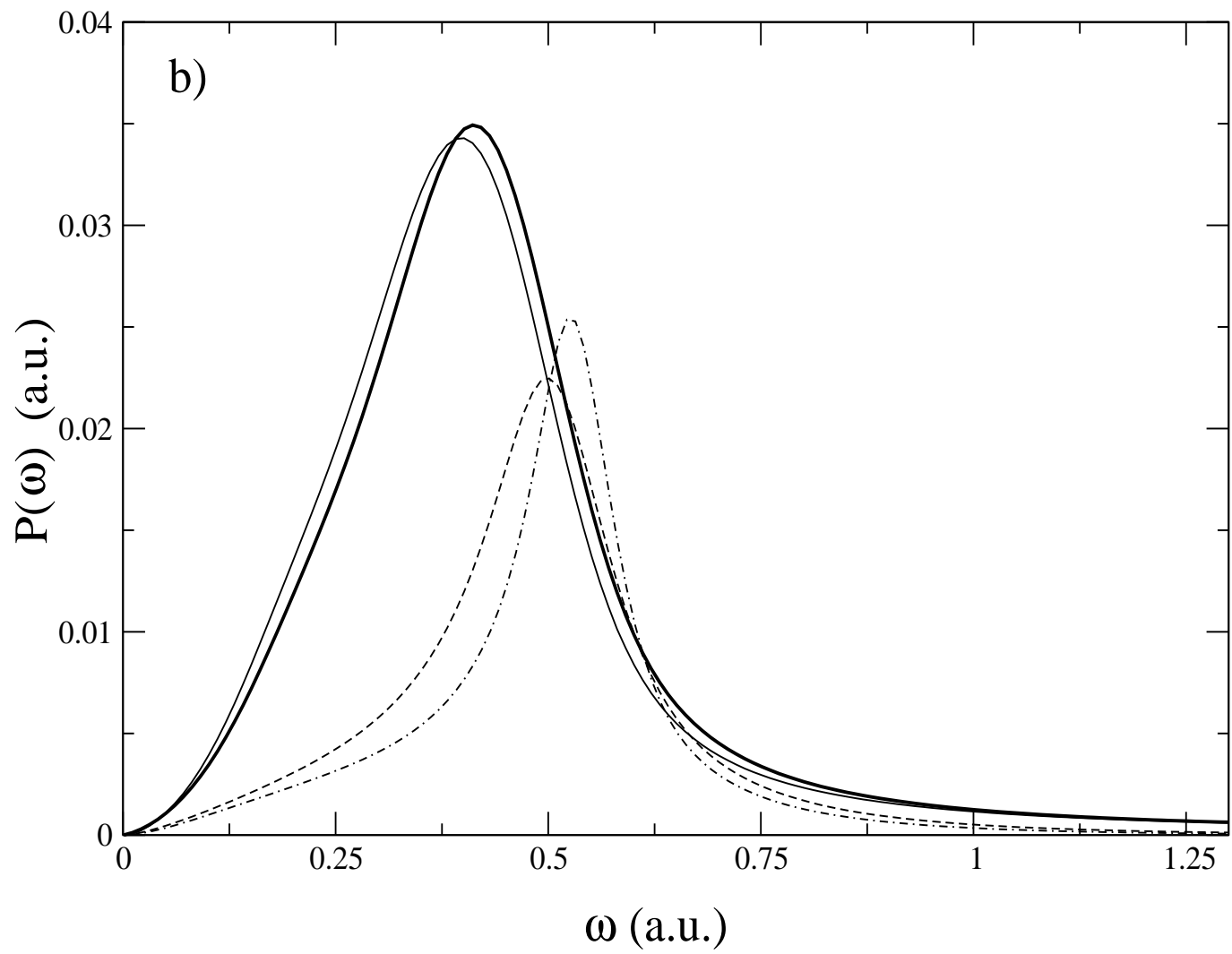
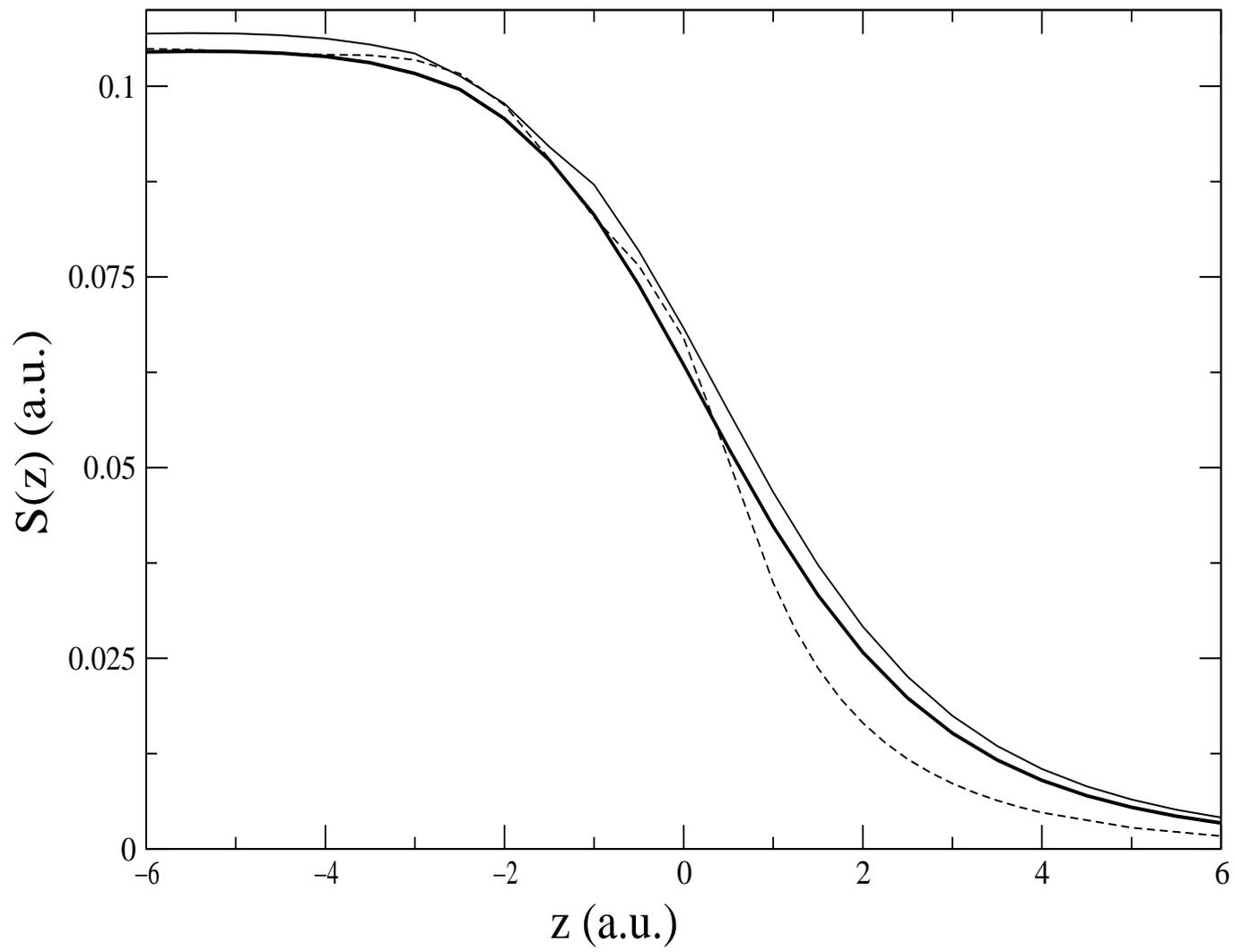
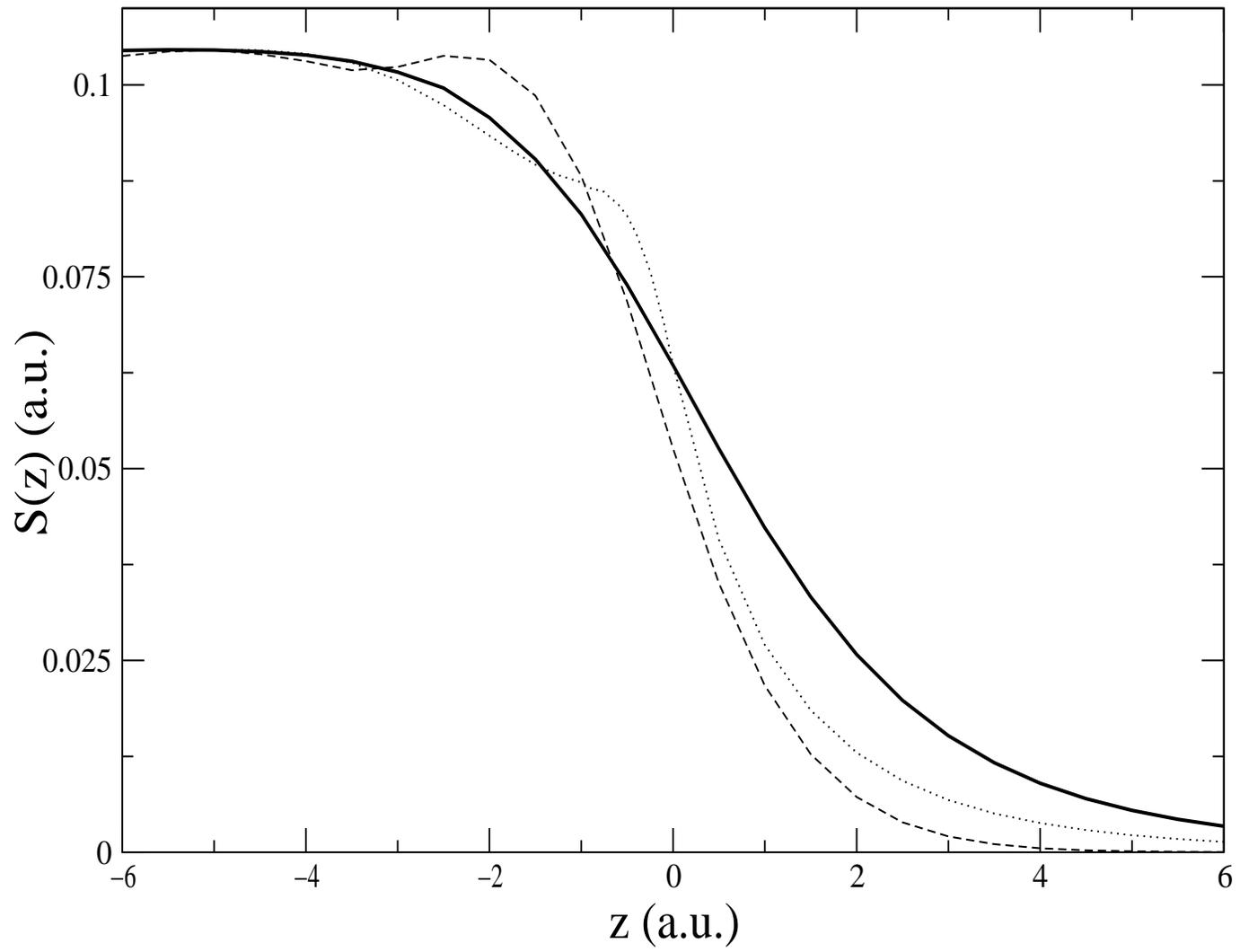
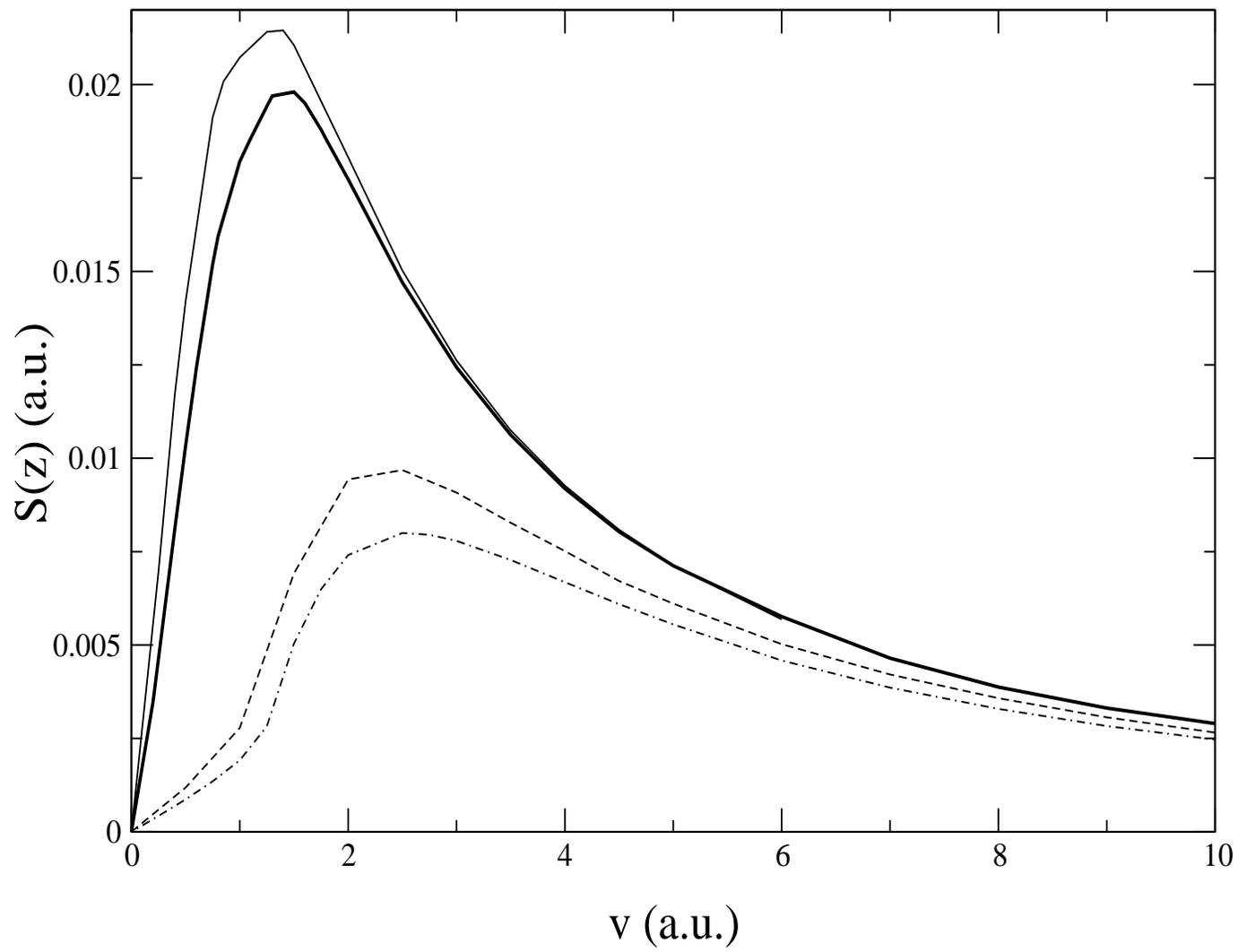
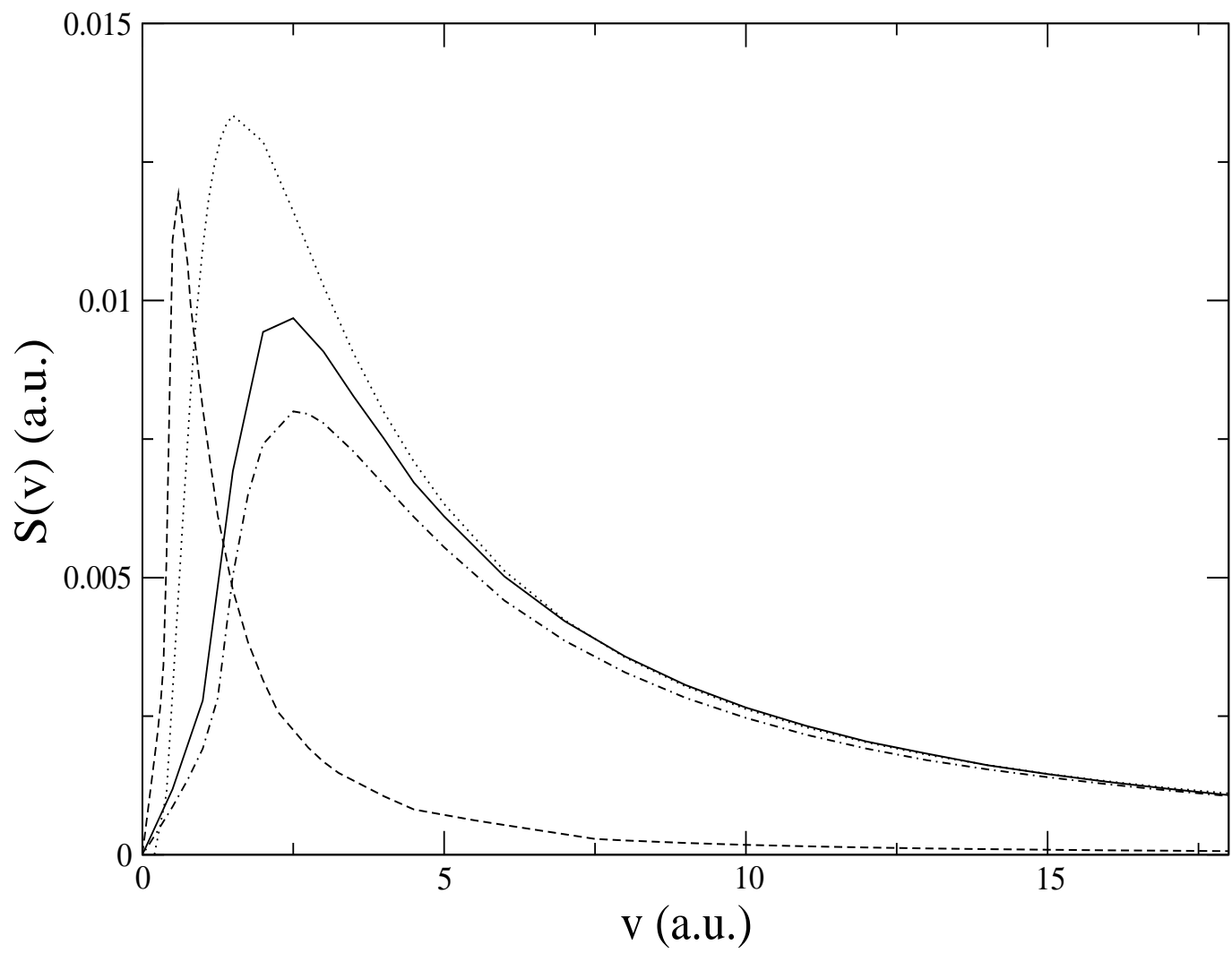
\end{document}